## Original article

# Ten years of research on ResearchGate: a scoping review using Google Scholar (2008–2017)

**Juan José Prieto-Gutiérrez**
*Complutense University of Madrid, Ciudad Universitaria S/N; jujpriet@ucm.es; ORCID 0000-0002-1730-8621*

**Abstract**
*Objective*: To analyse quantitatively the articles published during 2008–2017 about the academic social networking site ResearchGate.

*Methods*: A scoping bibliometric review of documents retrieved using Google Scholar was conducted, limited to publications that contained the word 'ResearchGate' in their title and were published from 2008 to 2017.

*Results*: The search yielded 159 documents, once a preliminary list of 386 documents retrieved from Google Scholar was filtered, which eliminated about 60% of the results that were bibliographic citations and not documents. Papers in journals were the most numerous type of documents (n = 73; 46%), followed by conference papers (n = 31; 19.5%). Contributing eight publications, two Spanish scholars (Delgado López-Cózar and Orduña-Malea, who were co-authors in each case) were the most prolific authors writing on this topic during the ten-year period. The keywords most used in the documents were 'ResearchGate' and 'Altmetrics'. The publications were cited frequently since 2014 (more than 90% of the total cites fell in that period), and those with more than one author were the most cited ones. The authors of the documents were mainly librarians and information science professionals, who wrote primarily as co-authors with colleagues from their own institutions, mostly published in English.

*Conclusions*: Interest in ResearchGate has grown since 2015, as evident from the number of articles published and the citations they received.

### Introduction

Social media are now being used extensively by academics for scholarly communication,[1] to share and disseminate[2] academic resources, to keep up with research, and to build professional networks[3] by establishing and strengthening relationships with other scholars, researchers, and professionals.[4] More and more members of the scientific community now access social networking sites as part of their daily work.[5]

ResearchGate is one such social network. Launched in 2008, it has become a point of reference for academics and scientists worldwide and, as of May 2018, had more than 15 million members (www.researchgate.net/about) and more than 100 million publications in its repository.

In participating in academic social networks, researchers themselves upload full texts of their documents. ReseachGate is increasingly used by scholars to upload the full text of their articles and make them freely available to everyone.[6] The site allows the publications to be tracked. Members can connect to others, engage in professional interaction, and register their interest in academic topics. Members also receive email alerts on activities related to their profile and publications. An analysis showed that ResearchGate is dominated by recent articles, which attract about three times as many views as older articles.[7] ResearchGate is a major source of full-text papers through Google Scholar.[8]

ResearchGate also offers a variety of statistics about members and institutions and provides recommendations on information relevant to research[9,10] in different disciplines.[11] The network has its own measure, namely RG Score: high ResearchGate scores are obtained primarily by asking and answering questions on the ResearchGate website. A high score could be used in recruiting for research positions, especially through the ResearchGate job vacancies service for academics,[12] or in evaluating grant proposals. The use of the platform also varies by levels of credibility[13] and the country of access.[14]

In short, ResearchGate has become one of the most widely used online academic social websites[15], bringing the benefits of online networking to an academic audience and publicizing the work of its members. Scholars in growing numbers are integrating social media tools into their communications.[16] In this context, it is important to ascertain the real interest the scientific community has in ResearchGate; to do so, the present paper assesses the coverage of documents published from 2008 to 2017 and directly related to ResearchGate, the academic social website, based on its first ten years of operation. To the best of our knowledge, no such analysis has been carried out of the type of publications produced, patterns of collaboration, gender of the member-authors, or the subject content or domain of the publications—and this study sought to gain some insights into the literature on ResearchGate through a bibliometric analysis of all ResearchGate-related publications retrieved using Google Scholar (GS).

### Methods

*Setting*
The data were collected in May 2019. The date range in GS was set to 2008–2017 (using the custom date range for each of these ten years and adding up the total).

*Database*
We used GS because it claims to be one of the largest scientific bibliographic databases in the world and is also a good alternative to retrieve journals that are not widely indexed and other documents such as blogs, presentations, articles in newspapers, working papers, and theses.

*Procedure*
The English version was used by manually entering the URL via http. Test queries were refined using the following command: allintitle:researchgate

Google Scholar compiles material from trusted academic websites (academic databases, universities, professional societies, repositories, libraries, etc) but also retrieves a greater number of citations and other non-academic documents such as patents. For this reason, the preliminary results were filtered. This possible lack of accuracy made it necessary to examine each of the records retrieved as a result of the search.

The filtering criterion was 'articles, excluding patents' to obtain the total number of items, sorted by the national or institutional web domain. The data were cleaned in GS to eliminate documents that were either duplicates or contained bibliographical errors. Finally, all the sets were sorted using the indicated bibliographic details.

Different fields of the records retrieved from GS were analysed: type of documents, authorship pattern and gender, country affiliation, keywords, and citations. For each year, the documents were categorized by the type of publication: articles, blog posts, conference papers, working papers, seminars and workshops, books, theses, and videos. The authors were ranked by the number of their contributions.

By examining the core list of records, we drew up a list of authors and extracted from that the names of 15 authors who had been the most productive over the ten-year period.

Keywords were compiled from a core list of scientific journals. Every keyword was counted and the keywords were ranked by frequency.

**Results**
The preliminary list comprised 386 records. We eliminated duplicates and corrected some errors related to citations and other details. The results of the first round of searches using GS for publications on the topic of ResearchGate for the years 2008–2017 were misleading. Although retrieved in response to the search term 'ResearchGate', these documents were not about the academic social network at all: most of them were papers on a variety of topics and were presented in PDF so that they could be uploaded on the ResearchGate website and therefore carried the term ResearchGate in the comments accompanying the files in PDF. Google Scholar thus failed to detect their irrelevance while retrieving the documents from repositories.

As can be seen from Table 1, the 386 records comprised 159 documents or publications (41.2%) and 227 citations (58.8%). There were no patents. Research papers in journals were the most frequent category and accounted for 46% of all the documents, followed by conference papers (19.5%), blogs (18.2%), seminars and workshops (6.3%), working papers (5.6%), books (2.5%), theses (1.25%), and videos (0.6%).

Table 1. Publications and citations on ResearchGate retrieved using Google Scholar by year and publication type (2008–2017)

| Year | Publication type | | | | | | | | | Citations N (%) |
|---|---|---|---|---|---|---|---|---|---|---|
| | Total records N (%) | Journal papers N (%) | Conference papers N (%) | Blogs N (%) | Seminars and workshops N (%) | Working papers N (%) | Books N (%) | Theses N (%) | Videos N (%) | |
| 2008 | 8 (2) | 1 (0.2) | 0 (0) | 0 (0) | 0 (0) | 0 (0) | 0 (0) | 0 (0) | 0 (0) | 7 (1.8) |
| 2009 | 12 (3.1) | 0 (0) | 0 (0) | 4 (1) | 0 (0) | 0 (0) | 0 (0) | 0 (0) | 0 (0) | 8 (2) |
| 2010 | 10 (2.5) | 0 (0) | 0 (0) | 0 (0) | 0 (0) | 1 (0.2) | 0 (0) | 0 (0) | 0 (0) | 9 (2.3) |
| 2011 | 10 (2.5) | 0 (0) | 0 (0) | 3 (0.8) | 0 (0) | 0 (0) | 0 (0) | 0 (0) | 0 (0) | 7 (1.8) |
| 2012 | 15 (3.8) | 0 (0) | 1 (0.2) | 2 (0.5) | 1 (0.3) | 0 (0) | 1 (0.25) | 0 (0) | 0 (0) | 10 (2.5) |
| 2013 | 19 (4.9) | 5 (1.3) | 1 (0.2) | 2 (0.5) | 0 (0) | 0 (0) | 0 (0) | 0 (0) | 0 (0) | 11 (2.8) |
| 2014 | 41 (10.6) | 4 (1) | 0 (0) | 3 (0.8) | 0 (0) | 1 (0.2) | 0 (0) | 0 (0) | 0 (0) | 33 (8.5) |
| 2015 | 78 (20.2) | 12 (3.1) | 13 (3.3) | 5 (1.3) | 0 (0) | 2 (0.5) | 1 (0.25) | 1 (0.25) | 1 (0.2) | 43 (11.1) |
| 2016 | 85 (22) | 14 (3.6) | 10 (2.5) | 3 (0.8) | 7 (1.7) | 4 (1) | 1 (0.25) | 0 (0) | 0 (0) | 46 (11.9) |
| 2017 | 108 (27.9) | 37 (9.5) | 6 (1.5) | 7 (1.8) | 2 (0.5) | 1 (0.2) | 1 (0.25) | 1 (0.25) | 0 (0) | 53 (13.7) |
| Total | 386 (100) | 73 (18.9) | 31 (8) | 29 (7.5) | 10 (2.5) | 9 (2.3) | 4 (1) | 2 (0.5) | 1 (0.2) | 227 (58.8) |

*Analysis of authors*
A total of 267 authors were identified, and Table 2 lists the 16 most productive of them; 16 authors had each contributed at least three papers and 18 had contributed at least two. The top two authors were Delgado López-Cózar and Orduña-Malea (each with 8 papers), both from Spain, followed by Rathemacher (6 papers) from USA. The gender balance was markedly skewed: 75% of the top authors were men.

Almost 75% of the articles were written by two or more authors, and the number of authors per paper increased over the last three years of the study; for productivity rankings, it may therefore be more appropriate to consider teams instead.

**Table 2.** Sixteen most productive authors of articles about ResearchGate as retrieved from Google Scholar and published during 2008–2017

| Author | Contributions | Gender | Country affiliation | Number of citations |
| --- | --- | --- | --- | --- |
| E Delgado López-Cózar | 8 | Male | Spain | 203 |
| A Orduña-Malea | 8 | Male | Spain | 203 |
| A Rathemacher | 6 | Female | USA | 28 |
| M Thelwall | 5 | Male | UK | 388 |
| A Martín-Martín | 5 | Male | Spain | 146 |
| J Lovett | 4 | Male | USA | 28 |
| M I Míguez-González | 4 | Female | Spain | 39 |
| A Dafonte-Gómez | 4 | Male | Spain | 39 |
| I Puentes-Rivera | 4 | Male | Spain | 39 |
| D He | 4 | Male | USA | 80 |
| Z Batooli | 3 | Female | Iran | 18 |
| K Kousha | 3 | Male | UK | 352 |
| W Jeng | 3 | Female | Taiwan | 77 |
| C Lutz | 3 | Male | Norway | 42 |
| C P Hoffmann | 3 | Male | Germany | 42 |
| I F Aguillo-Caño | 3 | Male | Spain | 5 |

*Analysis of keywords*
Keywords were collected from a core list of 73 scientific journals, which yielded a total of 243 keywords. Figure 1 presents the top keywords used by the 18 authors each of whom had contributed at least two papers.

The most used keywords were 'ResearchGate' (33) and 'Altmetrics' (14), closely followed by 'Bibliometrics' and 'Social media' (8). This observation was consistent with the fact that English was the most dominant language in the data set, because of the 18 most frequent keywords, 17 are English (the only non-English term, and the 10th most frequent, being the Portuguese term 'Ciência da Informação').

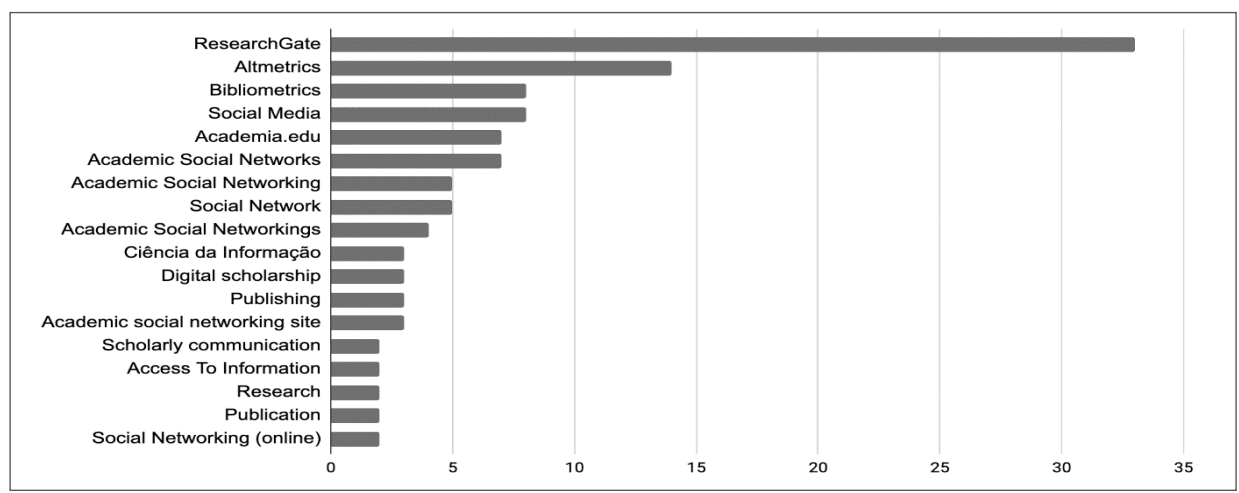

**Figure 1.** Most frequently used keywords in articles about ResearchGate as retrieved from Google Scholar and published during 2008–2017.

*Analysis of citations in Google Scholar*
All 159 publications together earned a total of 1465 citations. To make meaningful comparisons, the citations were analysed in terms of the type of publications being cited. For example, the 73 papers in journals were cited 1073 times. The score for the lowest quartile was 0, that is, 25% of the articles were not cited at all; median score or that for the second quartile was 3; and for the third quartile was 16. In other words, 75% of the papers earned 16 or fewer citations. 20 of these articles (27%) were not cited at all; 23 (31.5%) earned more than 10 citations and five (6.8%) more than 50 citations.

The most cited publication, cited 259 times, was 'ResearchGate: disseminating, communicating, and measuring scholarship?' by Thelwall and Kousha, published in 2015 in the *Journal of the Association for Information Science and Technology*.

The documents other than papers in journals (86 items, comprising blog posts, conference papers, working papers, seminars, books, theses, and videos) were cited 392 times.

Figure 2 compares the citations earned by each of the three groups of documents, namely all publications, papers in journals, and rest of the publications, over time. Citations to papers in journals show a marked rise since 2008.

During the first four years, 2008–2011, the total citations were close to zero. The numbers began rising in 2012 (60 citations in that year) and peaked at 450 in 2015. Until 2013, publications other than papers in journals were cited more often than papers in journals but the latter surged ahead from 10 to 90 in 2014 and earned 320 citations in 2015. In 2017, citations earned by papers in journals were practically the only category: the other types of documents were hardly cited at all in that year.

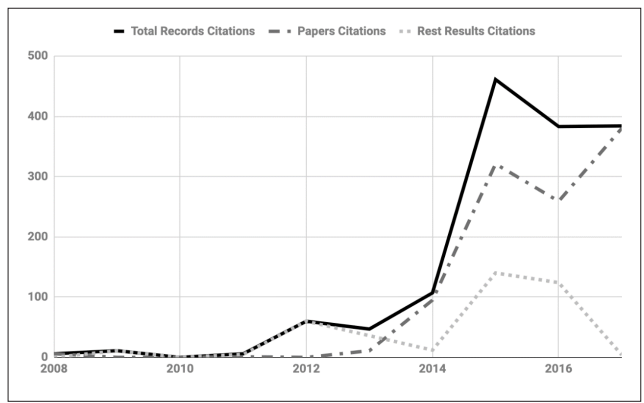

Figure 2. Citations earned by (1) all publications about ResearchGate, (2) papers in journals, and (3) other types of publications in Google Scholar from 2008 to 2017.

**Discussion**
Our initial search produced many irrelevant documents that were not about ResearchGate. Such errors can generate inaccurate results and show why it is important to comply with the policies of institutional repositories that promote open access. Once the irrelevant records were removed, we were left with 159 documents that qualified for detailed analysis and constituted the data set. Of these, the most numerous were papers in journals (46%), followed by conference papers (19.5%). These percentages probably reflect the relative preferences of the academic community for the two types of publications.

More than 84% of the documents were published between 2015 and 2017. The breadth of the GS database offers a large variety of different types of documents, although papers in journals represent the predominant category. This predominance demonstrates the maturity of the publications found on ResearchGate, as authors presumably prefer to offer their readers what they value most. The fact that our data set was so small seems to belie the strength of this academic network: the number of ResearchGate members runs to millions. However, we believe that it is only a matter of time; as mentioned earlier, research interest in ResearchGate has been growing steadily from 2015—90% of the documents and 80% of the citations in our data set date from that year.

The country as recorded in the affiliations of authors may be important in identifying a pool of authors working on a given topic, at least initially. In the present case, a large proportion (45%) of authors of papers about ResearchGate were affiliated to Spanish institutions, followed by those from USA (18%) and from the UK (12.5%).

The names in our data set, especially of the top 16 authors, required a great deal of standardization and validation of the data to retrieve the relevant articles. The gender gap was also significant but probably matches the smaller proportion of women in the international scientific community as a whole.

The analysis of author affiliation of the 15 most productive authors showed highly localized national networks, which, unfortunately, highlights the limited scope for international collaboration in researching the topic.

English continues to dominate as the language of scientific knowledge: 75% of the authors in our data set published more in English than in any other language and even non-English-speaking authors published in English, presumably to increase their international visibility, the number of citations being a measure of such visibility.

The choice of keywords reflects the respective content and research methods of the articles published in the ten years covered by the present paper and reveals priorities at the international level. Although 'ResearchGate' was bound to be the most frequent keyword for the present study and confirms ResearchGate's central position, 'academia.edu' was the 4th most frequent keyword, indicating that many authors had a keen interest in both the networks and have examined how the two compare. The second and the third most frequently used keywords were, respectively, 'Altmetrics' and 'Bibliometrics'. These measures of scholarly impact are increasingly used in academic social networks to measure various aspects of the documents hosted on the ResearchGate website. Other keywords, such as 'academic social network' and 'scholarly communication' and their variant forms (singular or plural) also underscore the predominance of English as the major language of communicating scientific knowledge.

The number of citations shows that documents on the academic social network ResearchGate are of interest to the academic community but particularly noteworthy is the shift in recent years in the number of citations in favour of papers in journals over other types of publications, a shift indicating that academic social networks are increasingly valued by the scientific community. The analysis of citations detected several highly cited documents, all of them being papers in journals. Citation is a social process and these articles by researchers in the field of library and information science, and in information technology, serve as references to other researchers.

**Conclusions**

A preliminary analysis of the number of publications on the topic of ResearchGate showed that research on the topic has been somewhat limited, especially given the millions who are members of that academic social network. During 2008–2017, 159 documents were published on the topic (as determined by searching Google Scholar). However, research interest in ResearchGate has been growing since 2015: 90% of the documents – all papers in journals – and 80% of the citations in our data set date from that year. The authors of the documents were mainly from the field of library and information science and had co-authored the papers with colleagues from their own institutions and had published mainly in English.